\documentclass[a4paper,10pt]{article}
\usepackage{graphicx}
\usepackage{fancyheadings}
\begin{document}
\noindent
{\it 11th Hel.A.S Conference}\\
\noindent
{\it Athens, 8-12 September, 2013}\\
\noindent
%
%
CONTRIBUTED POSTER\\
\noindent
\underline{~~~~~~~~~~~~~~~~~~~~~~~~~~~~~~~~
~~~~~~~~}
\vskip 1cm
%
%
\begin{center}
{\Large\bf
Updated analysis for the system V1464~Aql
}
\vskip 0.5cm
%
%
{\it
A. Liakos$^1$ and P. Niarchos$^2$
}\\
%
%
$^1$Institute for Astronomy, Astrophysics, Space Applications and Remote Sensing, National Observatory of Athens, I. Metaxa \& Vas. Pavlou St., Palaia Penteli, Athens, Hellas\\
$^2$Department of Astrophysics, Astronomy and Mechanics, Faculty of  Physics, National \& Kapodistrian University of Athens, Athens, Hellas
\end{center}
\vskip 0.7cm
%
%
\noindent
{\bf Abstract: }
New $BVRI$ light curves of the system V1464~Aql (ellipsoidal variable with a $\delta$~Scuti component) were obtained. Using these observations and taking into account previous studies of the system, its light curve model was reproduced and the pulsation frequencies of the $\delta$~Scuti component were recalculated. Moreover, the derived parameters were used to estimate the evolutionary status of both components.



\section{Introduction}
Generally, eclipsing binary systems (hereafter EBs) offer unique information for the calculation of stellar absolute parameters and the evolutionary status of stars. Especially, the cases of binaries with $\delta$~Sct components are extremely interesting, since they provide additional information (i.e. pulsation characteristics) for this part of the stellar lifetime. Therefore, the calculation of their absolute parameters and the identification of their oscillating characteristics help us to obtain useful conclusions for this 'unstable' part of stellar lifetime.

V1464~Aql ($m_{\rm{V}}$=8.98~mag, $P$=0.69777$^{\rm d}$) was observed spectroscopically by \cite{RU06}, who revealed that it is an eclipsing binary of F1-2 spectral type and measured the radial velocity of the primary component as $K_{1}$=31~(1)~km/s. The most detailed study of the system was made by \cite{DA13}, who found, using their own data, one pulsation frequency and calculated the absolute parameters of the system using the $q$-search method. In the present work, we use our new $BVRI$ light curves to create our model and to perform a detailed Fourier analysis in order to find the main characteristics of the pulsating component.

\section{Observations and analyses}
The system was observed during 13 nights in September of 2012 at the Athens University Observatory using a 40~cm Cassegrain telescope equipped with the CCD camera ST-10XME and the Bessell $BVRI$ photometric filters. A total of $\sim2000$ data points per filter (53.7~hrs in total) were collected in a time span of 16 days. Differential magnitudes were obtained with the software $Muniwin$ v.1.1.29 \cite{HR98}. The mean photometric error was 4.1, 3.3, 3.2 and 2.3~mmag for $B$, $V$, $R$ and $I$ filters, respectively.

\begin{figure}[h]
\centering
\begin{tabular}{ccc}
\includegraphics[width=5.5cm]{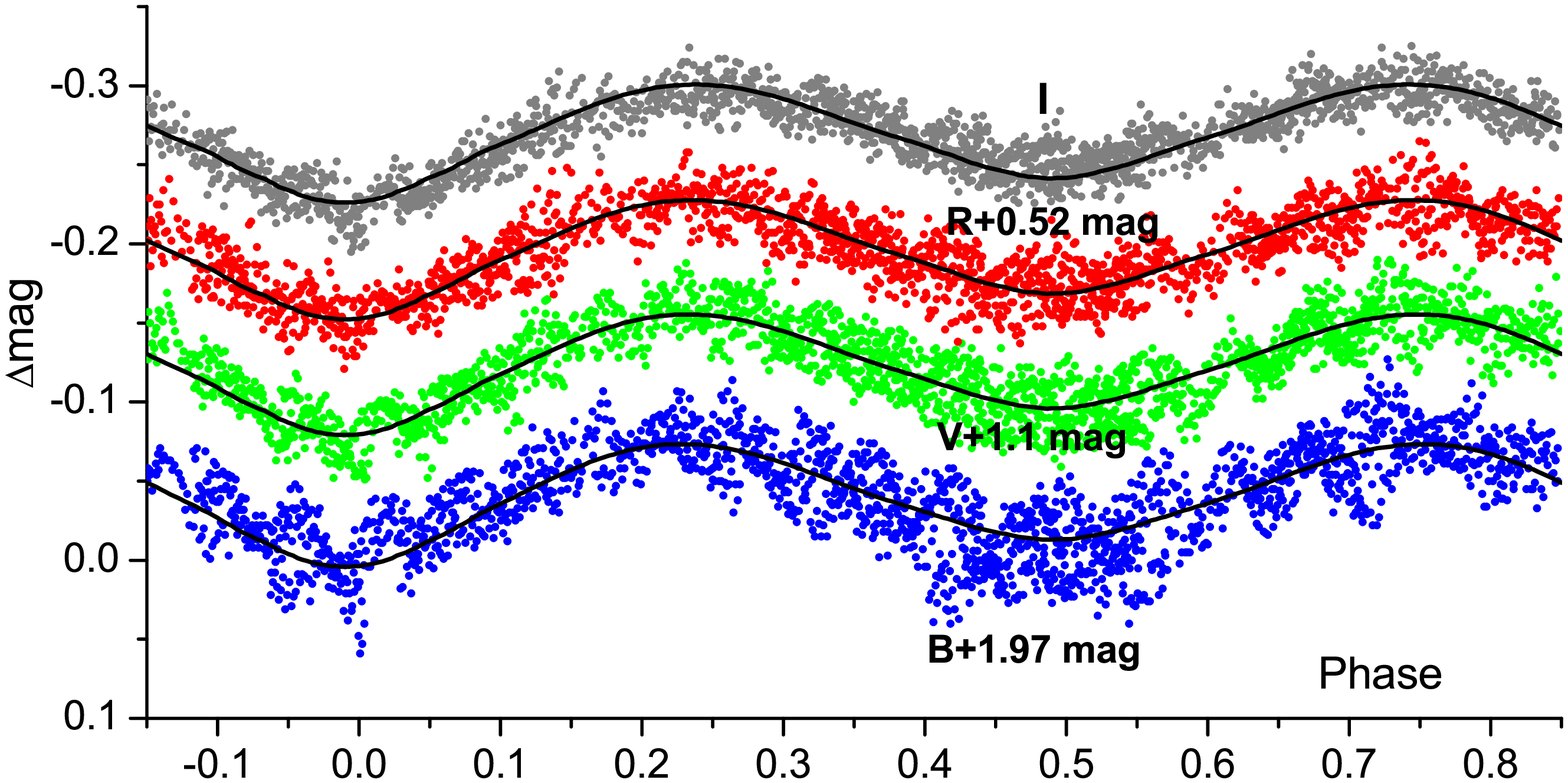}&\includegraphics[width=4.1cm]{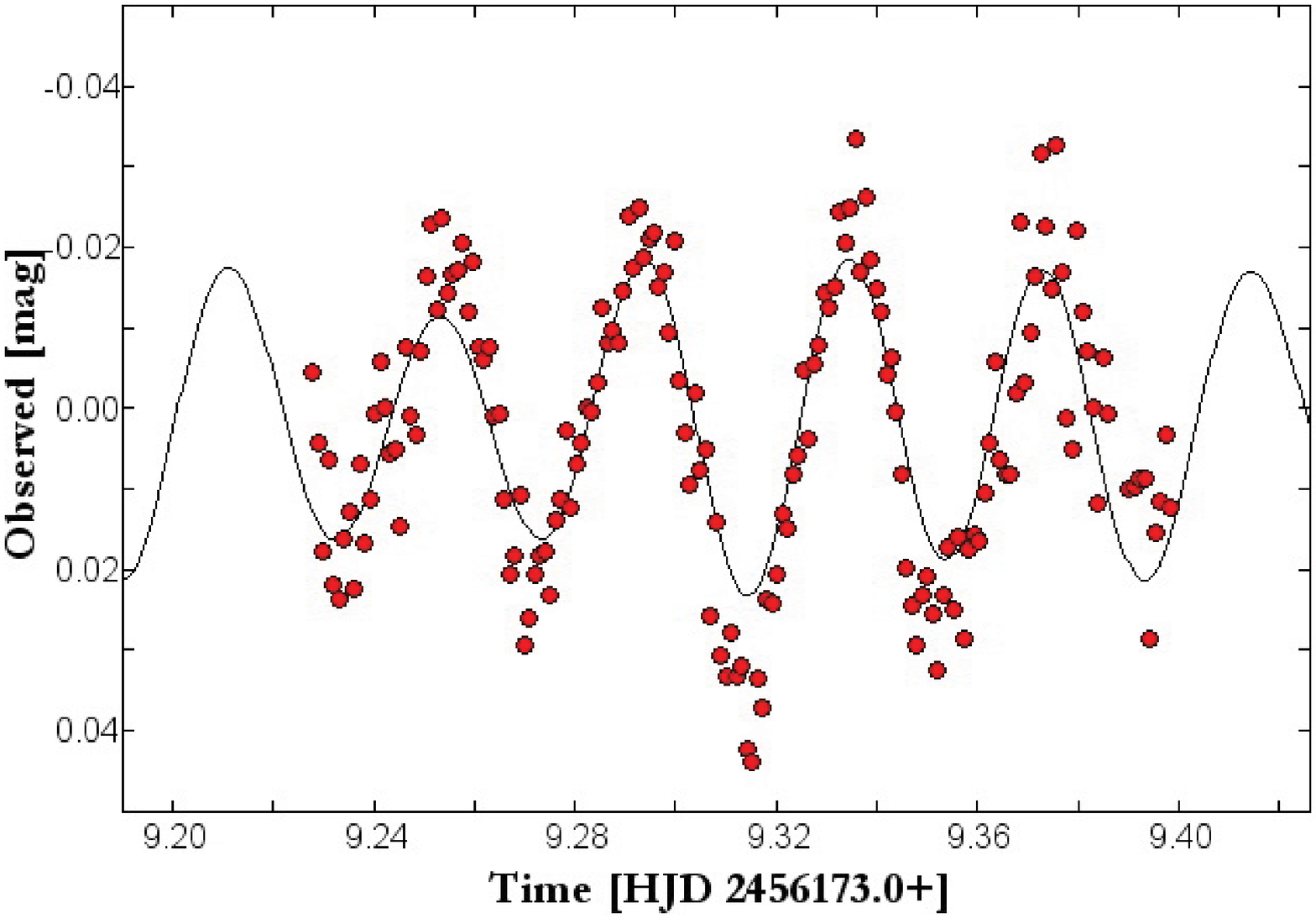}&\includegraphics[width=5.1cm]{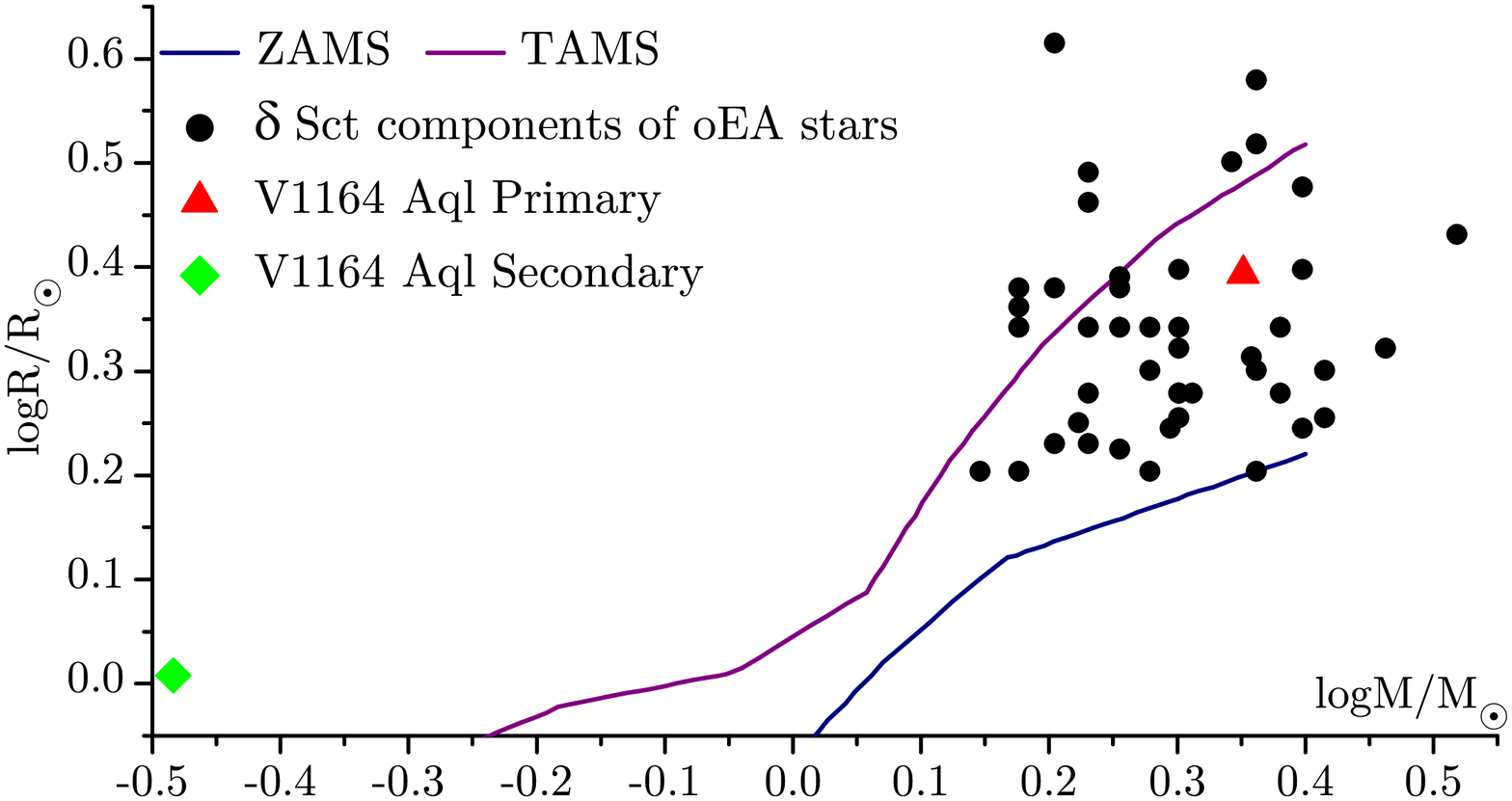}
\end{tabular}
\caption{Left: Observed (points) and synthetic (solid lines) LCs of V1464~Aql. Centre: Fourier fit on the longest data set in $B$-filter. Right: Positions of the components of V1464~Aql in the $M-R$ diagram together with other $\delta$~Sct stars-members of oEA stars (taken from \cite{LI12}).}
\label{Fig1}
\end{figure}

The light curves (hereafter LCs) have been analysed together with the primary component's radial velocity curve \cite{RU06} with the $Phoebe$ v.0.29d software \cite{PR05}. The temperature of the primary component was used as a fixed parameter ($T_{1}$=7100~K), based on the classification of \cite{RU06}. The $q$-search method was used in order to find the most probable “photometric” mass ratio of the system. The rest parameters were either given theoretical values or they were adjusted (for method and parameters details see \cite{LI11}). Finally, the best solution, using the $\chi^{2}$ criterion, was found in the conventional semi-detached mode. Synthetic and observed LCs are shown in Fig.\ref{Fig1}, while the derived model's parameters as well as the absolute parameters of the components are listed in Tab.\ref{Tab1}.

Frequency analysis was performed on the LCs residuals with the software $Period04$ v.1.2 \cite{LE05}(for method details see \cite{LI12}). The results showed that the primary component pulsates in a three frequency-mode, and they are given in Tab.\ref{Tab2}, where we list: frequency values $f$, semi-amplitudes $A$, phases $\Phi$ and S/N. Fourier fit on the longest data set is plotted in Fig.\ref{Fig1}.

\begin{table}[h]
\centering
\caption{Results of light curves analysis and absolute parameters of the components (comp.)}
\label{Tab1}
\scalebox{0.8}{

\begin{tabular}{lcc lcc cc l cc }
\hline
                             \multicolumn{8}{c}{\bf Light curve parameters}                        & \multicolumn{3}{c}{\bf Absolute parameters}   \\
\hline
\emph{Comp.:}       & \emph{P}&\emph{S}  &\emph{Filters:} &\emph{B} & \emph{V}& \emph{R} & \emph{I}&  \emph{Comp.:}         & \emph{P}&\emph{S}    \\
\hline
$T$ [K]             &   7100*  &5491(105)&   $x_{1}$      &  0.630  &  0.526  &   0.444  &  0.358  &$M$ [M$_{\odot}$]       & 2.2 (4) & 0.3 (1)    \\
$\Omega$            &  2.12 (4)& 2.07*   &   $x_{2}$      &  0.777  &  0.632  &   0.559  &  0.471  &$R$ [R$_{\odot}$]       & 2.5 (6) & 1.0 (2)    \\
$i$ [deg]&\multicolumn{2}{c}{46.8 (7)}   & $L_1/L_{\rm T}$&0.958 (4)&0.943 (3)&0.937 (3) &0.932 (2)&$L$ [L$_{\odot}$]       &  14 (1) & 0.8 (2)    \\
$q$      &\multicolumn{2}{c}{0.15 (1)}   & $L_2/L_{\rm T}$&0.042 (1)&0.057 (1)&0.064 (1) &0.068 (1)&$\alpha$ [R$_{\odot}$]& 0.6 (1) & 4 (1)        \\
\hline
\multicolumn{11}{l}{*assumed, $L_{\rm T}=L_1+L_2$, P=Primary, S=Secondary}
\end{tabular}}

\end{table}

\begin{table}[h]
\centering
\caption{Frequency search results for the pulsating component of V1464~Aql}
\label{Tab2}
\scalebox{0.58}{

\begin{tabular}{l cccc cccc cccc cccc }
\hline
        &       \multicolumn{4}{c}{$B$-filter}      &           \multicolumn{4}{c}{$V$-filter}      &           \multicolumn{4}{c}{$R$-filter}      &           \multicolumn{4}{c}{$I$-filter}      \\
\hline
	    &$f$ [c/d]	&$A$ [mmag]	&$\Phi$ [deg]&	S/N	&	 $f$ [c/d]	&$A$ [mmag]	&$\Phi$ [deg]&	S/N	&	$f$ [c/d]	&$A$ [mmag]	&$\Phi$ [deg]&	S/N	&	$f$ [c/d]	&$A$ [mmag]	&$\Phi$ [deg]&	S/N	 \\
\hline
$f_{1}$	&24.621 (1)	&	17.7 (3)&	341 (1)	 &	21.9&	24.621 (1)	&	14.3 (3)&	341 (1)	 &	22.6&	24.619 (1)	&	11.1 (3)&	347 (2)	 &	15.7&	24.621 (1)	&	8.6 (2)	&	339 (2)	 &	20.6 \\
$f_{2}$	&14.609 (4)	&	2.7 (3)	&	43 (6)	 &	3.8	&	14.585 (3)	&	2.9 (3)	&	99 (5)	 &	5.3	&	14.603 (5)	&	2.0 (3)	&	65 (9)	 &	3.5	&	14.601 (6)	&	1.4 (2)	&	66 (10)	 &	3.0	 \\
$f_{3}$	&29.621 (4)	&	2.6 (3)	&	83 (7)	 &	4.0	&	29.627 (4)	&	2.5 (3)	&	80 (6)	 &	5.0	&	29.640 (6)	&	1.8 (3)	&	84 (10)	 &	3.2	&	29.630 (5)	&	1.5 (2)	&	78 (9)	 &	3.2	 \\
\hline
\end{tabular}}

\end{table}

\section{Discussion and conclusions}
We analysed new multicolour LCs of V1464~Aql. New results for the absolute elements are derived and new pulsation frequencies were detected. The differences between our results and those of \cite{DA13}, regarding the evolutionary status of the components, probably come from the $q$-search method. More detailed spectroscopic observations are needed for final conclusions.

Two more pulsation frequencies for the primary component were found in comparison with the work of \cite{DA13} increasing the total detected frequency number to 3. In $M-R$ diagram (Fig.\ref{Fig1}) it is shown that the present results for the pulsating component of V1464~Aql are compatible with other for $\delta$~Scuti stars in binaries.


\noindent

\end{document}